\documentclass[preprint2]{aastex6}

\hypersetup{linkcolor=red,citecolor=blue,filecolor=cyan,urlcolor=magenta}

\received{2017 April 28}\revised{2017 June 18}\accepted{2017 July 1}\published{2017 August 4}

\shorttitle{Midterm Periodicity of the Mount Willson magnetic index}
\shortauthors{Feng et al.}

\begin{document}

\title{Midterm Periodicity Analysis of the Mount Wilson Magnetic Indices Using the Synchrosqueezing Transform}

\author{
Song FENG\altaffilmark{1,2}, Lan YU\altaffilmark{3}, Feng WANG\altaffilmark{1}, Hui DENG\altaffilmark{1}, Yunfei YANG\altaffilmark{1}
}
\altaffiltext{1}{Yunnan Key Laboratory of Computer Technology Application / Faculty of Information Engineering and Automation, Kunming University of Science and Technology, Kunming 650500, China; {\it feng.song@astrolab.cn}}
\altaffiltext{2}{CAS Key Laboratory of Solar Activity, National Astronomical Observatories, Beijing 100012, China}
\altaffiltext{3}{Department of mechanical and electrical engineering, Yunnan Land and Resources Vocational College, Kunming 650217, China}



\begin{abstract}
A novel time--frequency technique, called the synchrosqueezing transform (SST), is used to investigate the midterm periodic variations of magnetic fields on the solar surface. The Magnetic Plage Strength Index (MPSI) and the Mount Wilson Sunspot Index (MWSI), measured daily by the Mount Wilson Observatory between 1970 January 19 and 2012 January 22, are selected. The short-, mid, and longer-term periodicities are represented and decomposed by the SST with hardly any mode mixing. This demonstrates that the SST is a useful time--frequency analysis technique to characterize the periodic modes of helioseismic data. Apart from the fundamental modes of the annual periodicity, $\sim$27 day rotational cycle and $\sim$11 year solar cycle, the SST reveals several midterm periodicities in the two magnetic activity indices, specifically, $\sim$157 day (i.e., Rieger-type periodicity), and $\sim$1.3 and 1.7 years. The periodic modes, with 116.4 and 276.2 day periodicity in the MPSI, with 108.5 and 251.6 day periodicity in the MWSI, and the 157.7 day periodicity in the two indices, are in better accord with those significant periodicities derived from the Rossby waves theoretical model. This study suggests that the modes are caused by the Rossby waves. For the 1.30 and 1.71 year periodicity of the MPSI, and the 1.33 and 1.67 year periodicity of the MWSI, our analysis infers that they are related to those periodicity with the same timescale in the interior of the Sun and in the high atmospheric layers. 
\end{abstract}

\keywords{Sun: activity --- Sun: magnetic fields --- Sun: rotation}
\maketitle


\section{Introduction} \label{sec:intro}

The quasi-periodic oscillations of solar activity indicators, such as the numbers and areas of sunspots, coronal indices, plage areas, and soft X-flare indices, exhibit wide variations from days to decades \citep{2000ApJ...540.1102L,2002SoPh..205..177R,2003ApJ...591..406B,2008MNRAS.386.2278D,2009MNRAS.392.1159C}, with the most famous periodic modes being the rotational cycle with a 27 day period and the solar cycle with an 11 year period. The former clearly correlates with the monthly solar rotation and the latter with the polarity reversal of magnetic fields \citep{2015JASTP.122...18D}. A large number of works have been demonstrated that the periodic regions,  i.e., those between 27 days and 11 years, called midterm periodicities, still exist with different timescales in many modes \citep{2003ApJ...591..406B}. 

A lot of studies have been reported that the 1.3 and 1.7 year periodicities indeed exist in the solar interior, the atmospheres of the Sun, and the interplanetary space \citep[e.g.,][]{2000Sci...287.2456H,2012JASTP..89...48S,2014JGRA..119.9357Z,2002SoPh..205..165K,2013ApJ...778...28C}. Furthermore, they are associated with the basic process of solar dynamos, and can be clearly observed in the Sun. Recently, some authors \citep[e.g.,][]{2014SoPh..289..707C,2015JASTP.122...18D} have also suggested that coupled periodicities exist among the Sun, the interplanetary space, and the Earth's magnetosphere.

The Rieger-type periodicity of 154 days was first found \citep{1984Natur.312..623R} in the $\gamma$-ray flares. Subsequently, it has been found in some other indices, including the solar electron flux, H$\alpha$ flare activity, and radio flux at 2800 MHz \citep{2002JGRA..107.1318B,2002A&A...394..701K,2006MNRAS.373.1577C}. However, its physical mechanism has not been fully understood thus far. 

Other periodicities, from several to hundreds of days---for example, 13.5, 120, and 276 days---as well as the annual period, have also been determined \citep{2002SoPh..205..177R,2003ApJ...591..406B,2014JGRA..119.9357Z,2015MNRAS.451.4360K,2015AJ....150..171X,2016AJ....151...70D,2013ApJ...778...28C}.

Investigating the midterm periodic modes of solar activity indices, especially those of Rieger type, and the 1.3 and 1.7 year ones, can provide further  crucial knowledge to clarify the temporal evolutions of solar magnetic fields and to understand the basic process of solar dynamo models. In this study, we use two magnetic indices measured by the Mount Wilson Observatory (MWO) to investigate the midterm periodic variations of solar magnetic activities. These two magnetic indices are the Magnetic Plage Strength Index (MPSI) and the Mount Wilson Sunspot Index \citep[MWSI;][]{1991SoPh..135..211U,1991AdSpR..11..217U}. 

Fourier and wavelet transforms are usually used to analyze the frequency modes of solar quasi-periodic oscillations. These two techniques can only characterize the spectra and are unable to decompose them. So, the empirical mode decomposition (EMD) that decomposes a given signal into a set of intrinsic mode functions (IMFs) is used to analyze periodic variations of solar activities. Through the extraction of IMFs, the EMD can capture the non-stationary features of a quasi-periodic signal. Furthermore, it has been proven to be a good tool for analyzing solar oscillations \citep[e.g.,][]{2012MNRAS.423.3584L,2013Ap&SS.343...27D,2015MNRAS.451.4360K,2016ApJ...830..140G}. But some features of the EMD, like mode mixing and splitting, and aliasing, restrict its applications \citep{mandic2013empirical} . 

\begin{figure}
	\centering
        \includegraphics[width=8.5cm]{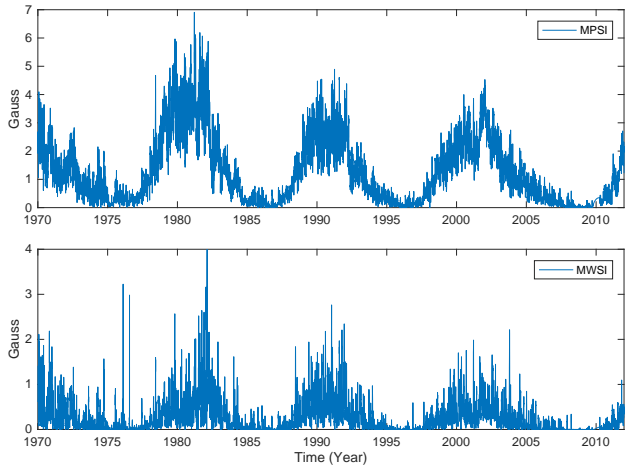}
\caption{Distributions of the MPSI and the MWS measured from the Mount Wilson Observatory. The data is measured from 1970 January 19 to 2012 January 22.}
		\label{fig1}
\end{figure}

\cite{daubechies2011synchrosqueezed} proposed a novel time--frequency analysis technique, called the synchrosqueezing transform (SST), which resembles the EMD but with a firm theoretical foundation. The SST not only can precisely represent the spectra of a given signal, but also reconstruct its intrinsic modes under bounded perturbations and Gaussian white noise \citep{thakur2013synchrosqueezing}. Therefore, we use this technique to decompose the periodic modes of the two solar magnetic indices. 

This paper is structured as follows. Section 2 describes the data sets of the MPSI and the MWSI. Section 3 explains the time--frequency representation and decomposition procedures based on the SST. The periodic modes of the MPSI and the MWSI are given and discussed in Section 4. Finally, the conclusions are presented in Section 5.

\section{Data sets and Data reduction}

The MPSI and the MWSI are calculated using the daily magnetogram to reflect solar magnetic field variations. The MPSI is determined from the magnetic field strength of the regions that are the plage/facular regions and the regions outside of sunspots on the whole disk. In thees regions, the absolute values of the magnetic field strengths from 10 to 100 G are summed and divided by all pixels of the magnetogram. Thus, the MPSI can be considered to represent the solar weak magnetic field activity. The MWSI values are determined in much the same manner as the MPSI, but only those pixels whose absolute vlaues are greater than 100 G are considered.

The selected MPSI and MWSI values have been recorded in every day from1970 January 19 to 2012 January 22. In this duration, only 11,084 pointd were recorded out of the total of 15,344 days, and others failed. So, we used a linear interpolation method to fill in those unrecorded data. Figure 1 shows the interpolated MPSI and WMSI data.  

\section{Method and Mode Extraction} \label{sec:sst}

The SST is a time--frequency analysis technique based on a continuous wavelet using a spectral reassignment method, and concentrates the frequency content around the instantaneous frequencies in the wavelet domain. A continuous wavelet transform $W_s$ of a signal $s(t)$ is defined by

\begin{equation}
W_s(a,\tau)=\frac{1}{\sqrt a}\int^{\infty}_{-\infty}s(t)\psi^*(\frac{t-\tau}{a})dt,
\end{equation}
where $\psi$ is a mother wavelet that can be shifted in time $\tau$ and stretched by the scale $a$, and $\psi^*$ denotes its complex conjugate. The fundamental principle of the SST is to concentrate the energy in the time--frequency map into the instantaneous frequency to decrease spectral smearing. The instantaneous frequency $\omega_s(a,\tau)$ is calculated using the derivative of the wavelet transform at any point $(a, \tau)$ with respect to $\tau$, 

\begin{equation}
\omega_s(a,\tau)=\frac{-i}{2\pi W_s(a,\tau)}\frac{\partial W_s(a,\tau)}{\partial \tau}.
\end{equation}

\begin{figure}
	\centering
        \includegraphics[width=8.5cm]{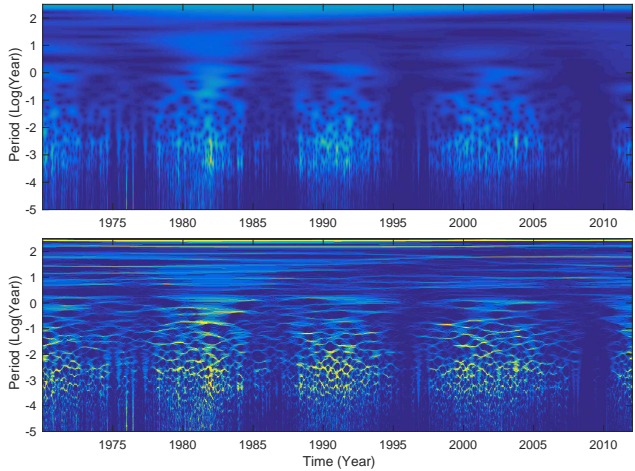}
\caption{Spectrum derived from the continuous wavelet transform and the SST. The upper and bottom panels are the spectrum of the continuous wavelet and the SST, respectively. The y-axis represents the period in a logarithmic scale for better visualization. Compare tdo the continuous wavelet spectrum, the spectrum represented by the SST is much sharper.}
		\label{fig2}
\end{figure}

The instantaneous frequencies $\omega_s(a,\tau)$ are the ridges in the time--scale plane. Subsequently, the frequencies around these ridges are squeezed to decrease spectra smearing. So, the information in the time--scale domain is converted into the time--frequency domain, i.e., every point $(a, \tau)$ is transferred to $(\omega_s(a,\tau),\tau)$. The process is called synchrosqueezing. Since the time $\tau$ and the scale $a$ are discrete values, $W_s(a,\tau)$ is calculated at the $a_k$, with a scale step $(\Delta a)_k=a_{k}-a_{k-1}$. Thus, the SST $T_s(\omega,\tau)$ is determined only at the centers $\omega_\ell$ of the frequency bins $[\omega_{\ell}- \Delta \omega/2,\omega_{\ell}+\Delta \omega/2]$, with $\Delta\omega=\omega_{\ell}-\omega_{\ell-1}$:

\begin{equation}
T_s(\omega_\ell , \tau)=\frac{1}{\Delta\omega}\sum_{a_k} W_s(a_k,\tau)a_k^{-3/2}(\Delta a)_k,
\end{equation}
where $a_k$ is bounded by $|\omega(a_k,\tau)-\omega _{\ell}|\leqslant \Delta\omega/2$.
This means that the frequency band that is less than half of the bandwidth around the centers $\omega_\ell$ is used to concentrate the frequency information. Thus, the time--frequency representation $T_s(\omega_\ell , \tau)$ is synchrosqueezed along the frequency or scale axis \citep{tary2014spectral}. The fully discretized estimate of $T_s(\omega_\ell,\tau)$ is denoted by $\tilde{T}_{\tilde{s}}(\omega_{\ell},t_m)$, where $t_m$ is the discrete sampling localization in time domain.

\begin{figure*}
	\centering
        \includegraphics[width=14cm]{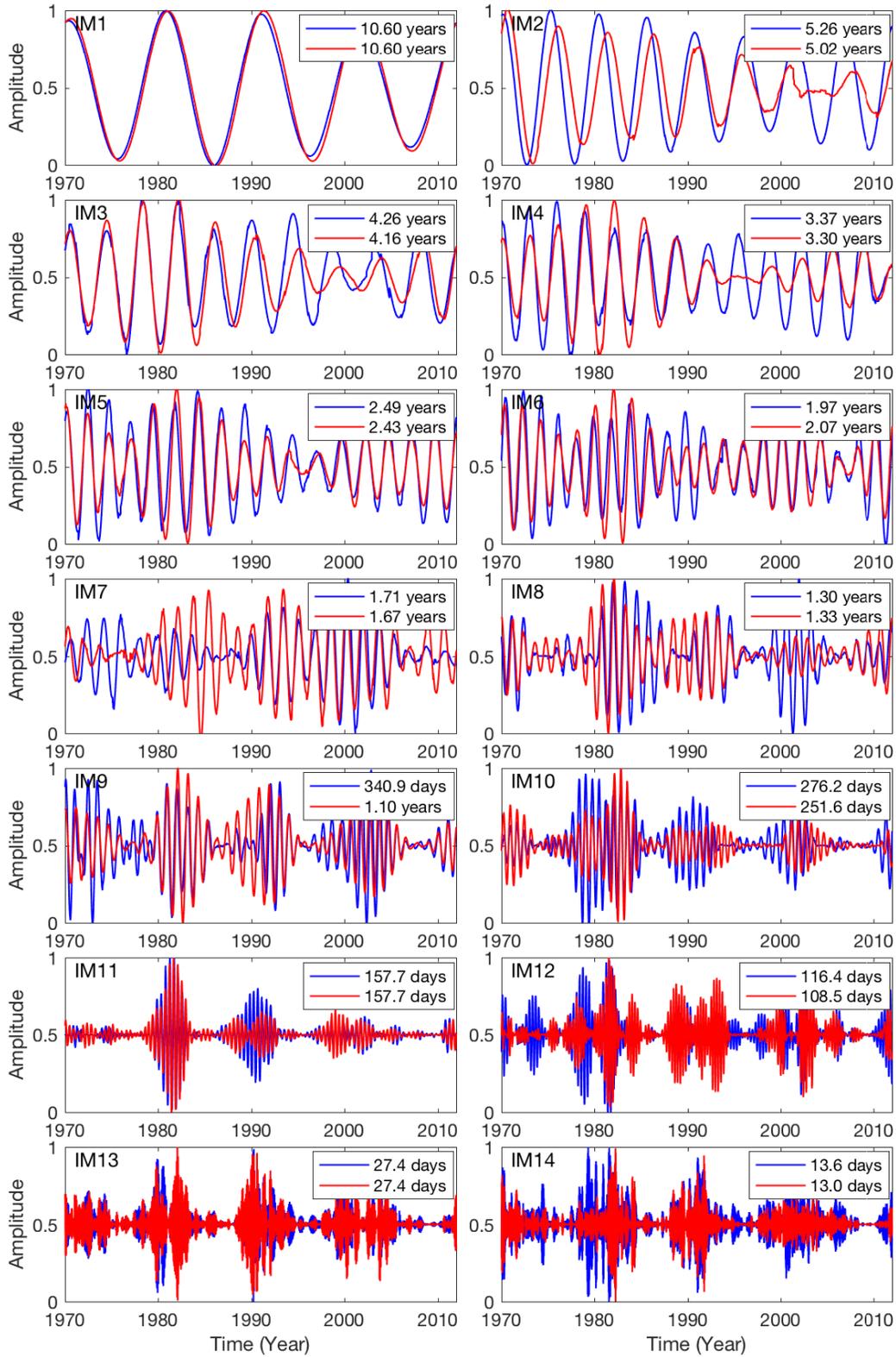}
         \caption{Periodic modes of the MPSI and the MWSI decomposed and reconstructed using the SST. To better compare the difference between the indices, the modes with the same number are superposed on a subplot. The modes of the MPSI are plotted in blue, and those of the MWSI in red. Their corresponding periodicities are given in the legend of each subplot. It notes that the amplitudes of all modes are normalized to their maximum values for better comparison. The amplitude of each mode is listed in Table 1 in the form of the mean strength ratio.}
        \label{fig3}
\end{figure*}

Each component $s_k$ of the original signal $s(t)$ is recovered from the discrete SST $\tilde{T}_{\tilde{s}}$ by inverting the continuous wavelet.

\begin{equation}
s_k(t_m)=2C_{\psi}^{-1}\Re \left(\sum_{\ell\in L_k(t_m)}\tilde{T}_{\tilde{s}}(w_\ell,t_m)\right),
\end{equation}
where, $C_\psi$ is a normalization constant from the selected wavelet and $\Re$ denotes the real part of the discrete SST. The individual component is reconstructed in the small frequency band $\ell\in L_k(t_m)$ around the $k$th component. The step resembles a filter on the time--frequency domain. Here, the frequency components can be estimated by a penalized forward--backward greedy ridge extraction method. A more detailed description of the SST can be found in \cite{daubechies2011synchrosqueezed} and \cite{thakur2013synchrosqueezing}.

\begin{table*}
\centering  
\caption{Periodic modes of the MPSI and the MWSI decomposed using the SST}
\begin{tabular}{ccccc}
\hline\hline
Mode &\multicolumn{2}{c}{MPSI} &  \multicolumn{2}{c}{MWSI}\\
 &Period& Mean Strength Ratio\tablenotemark{a}  &  Period& Mean Strength Ratio\tablenotemark{a} \\
\hline
1	&10.60$^{+0.50}_{-0.48}$	\ years		&1.00&	10.60$^{+0.50}_{-0.48}$	\ years		&	1.00\\
2	&5.26$^{+0.25}_{-0.24}$		\ years		&0.18&	5.02$^{+0.24}_{-0.23}$	\ years		&	0.14\\
3	&4.26$^{+0.20}_{-0.19}$		\ years		&0.05&	4.16$^{+0.20}_{-0.19}$	\ years		&	0.15\\
4	&3.37$^{+0.16}_{-0.15}$		\ years		&0.10&	3.30$^{+0.16}_{-0.15}$	\ years		&	0.13\\
5	&2.50$^{+0.12}_{-0.11}$		\ years		&0.10&	2.43$^{+0.12}_{-0.11}$	\ years		&	0.15\\
6	&1.97$^{+0.09}_{-0.09}$		\ years		&0.06&	2.07$^{+0.10}_{-0.09}$	\ years		&	0.14\\
7	&1.71$^{+0.08}_{-0.08}$		\ years		&0.05&	1.67$^{+0.08}_{-0.08}$	\ years		&	0.06\\
8	&1.30$^{+0.06}_{-0.06}$		\ years		&0.03&	1.33$^{+0.06}_{-0.06}$	\ years		&	0.08\\
9	&340.9$^{+16.3}_{-15.6}$	\ days		&0.10&	1.10$^{+0.05}_{-0.05}$	\ years		&	0.20\\
10	&276.2$^{+13.2}_{-12.6}$	\ days		&0.05&	251.6$^{+12.0}_{-11.5}$	\ days		&	0.16\\
11	&157.7$^{+7.5}_{-7.2}$		\ days		&0.04&	157.7$^{+7.5}_{-7.2}$	\ days		&	0.16\\
12	&116.4$^{+5.6}_{-5.3}$		\ days		&0.04&	108.5$^{+5.2}_{-4.9}$	\ days		&	0.20\\
13	&27.4$^{+1.3}_{-1.2}$		\ days		&0.15&	27.4$^{+1.3}_{-1.2}$	\ days		&	0.33\\
14	&13.6$^{+0.6}_{-0.6}$		\ days		&0.10&	13.0$^{+0.6}_{-0.6}$	\ days		&	0.25\\
\hline
\end{tabular}
\tablenotetext{a}{The ratio between the mean amplitude of each mode and that of the 11 year mode in the same index.}
\end{table*}

Figure 2 shows the frequency content of the MPSI based on the continuous wavelet transform (upper panel) and on the SST (bottom panel). The y-axis uses a logarithmic scale for better visualization. Compared to the spectra of the continuous wavelet, the spectra of the SST are obviously sharper around those intrinsic modes. The time--frequency ridges will be extracted to reconstruct their periodic modes. Here, the bump wavelet is selected, and the bin $\Delta\omega$ is set to 4. The errors of each mode are estimated with the bin values. 

A total of 14 modes are decomposed by the SST and are shown in Figure 3. To assess the relation between the modes from the two magnetic indices, the modes with the same number are superposed onto a subplot. The periodic modes of the MPSI are plotted in blue, and those of the MWSI in red. The legend of each subplot indicates the periodic values of each curves. It notes that their amplitudes are normalized to their maximum values for better comparison. 

Table 1 lists the periodic values and their errors, and their mean strength ratio between the mean amplitude of each mode and the 10.6 year (IM1) amplitude; the aim is to visually display the relative strengths among the modes. 

The frequency spectra of each mode, based on the SST, in the MWSI are shown in Figure 4 to evaluate whether or not the reconstructed modes exist mode mixing. Due to the similar results, the spectra of those modes in the MPSI do not show there. As shown in Figure 4, the frequency content of each mode almost only exists in a narrow period band, the center of which is the localization of the intrinsic mode, and hardly any in any other regions. This demonstrates that the SST technique can locate the frequency components with a high spectrum resolution,  and meanwhile, decompose and reconstruct them with hardly any mode mixing.

\section{Result and Discussion} 
As listed in Table 1, the shortest mode (IM14) is the $\sim$13 day periodicity. Apart from the fundamental periodicities such as $\sim$27 day (IM13) and $\sim$11 year (IM1) ones, some midterm periodicities can also be characterized with the SST, such as the $\sim$157 day, and $\sim$1.3 and $\sim$1.7 years (IM11, IM8, and IM7) modes in the two indices. 

From the mean strength ratio listed in Table 1, one can find that the relative large amplitude are in the modes with 10.6 years (IM1), 5.26 years (IM2) and 27.4 days (IM13) in the MPSI, whereas these are in the modes with 10.6 years (IM1), 27.4  and 13.0 days (IM13 and IM14) in the MWSI. The amplitudes of the other modes are relatively small. The 13.6 day periodicity in the MPSI and the 13.0 day periodicity in the MWSI can be considered as a subharmonic of the 27 day rotation cycle \citep{2003ApJ...591..406B}. Similarly, the 5.26 year (MPSI) and 5.02 year (MWSI) timescales in IM2 are approximately half of the 10.6 year timescale (IM1). Thus, the relatively large amplitudes are the periodicities of $\sim$11 years and $\sim$27 days, demonstrating that the two periodic signals are a significant mode.

\begin{figure*}[t]
	\centering
        \includegraphics[width=14cm]{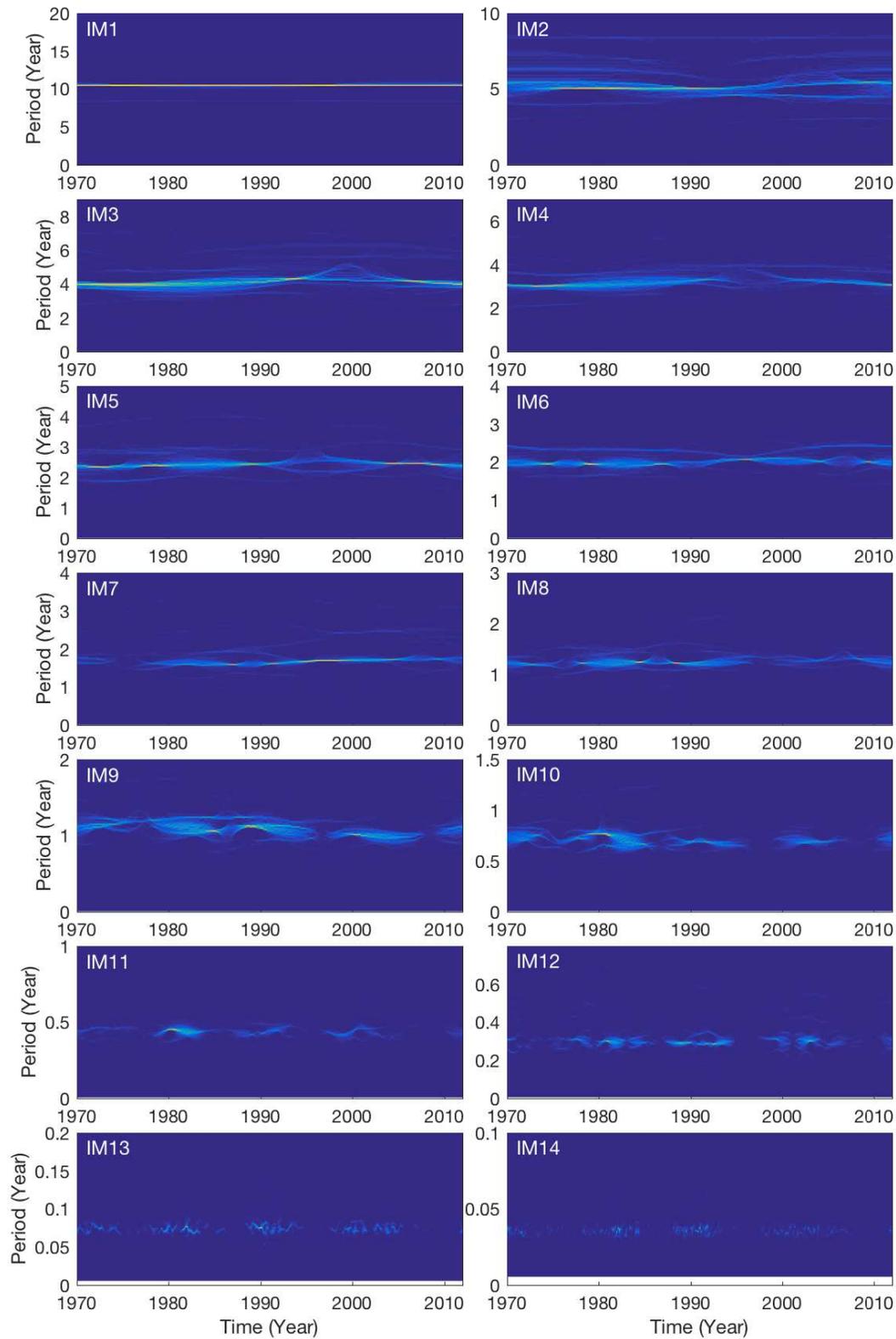}
        \caption{Time--frequency representation of each mode based on the SST in the MWSI. The frequency content of each mode only exists in a narrow frequency band, the center of which is the localization of each mode.}
        \label{fig4}
\end{figure*}
 
The periodicities of 340.9 days (IM9), 1.97 years (IM6), and 4.26 years (IM3) in the MPSI, and those of 1.10 years (IM9), 2.07 years (IM6) and 4.16 years (IM3) in the MWSI are inferred to be the first, second, and fourth harmonics of the annual periodicity signal. \cite{2000Sci...287.2456H} found the annual periodicity from the helioseismic probing of the solar interior only at high latitudes, and argued against the mode generated by the annual systematic observational errors. However, it is difficult to quantify the possibility that this is not due to the influence of seasonal variations, although the periodicity has been found in several solar activity indices \citep{2009SoPh..257...61J,2012ApJ...747..135L}. From the extracted annual periodicities, we tend to infer that the periodicities are possibly caused by the Earth's heliolatitude. It must be pointed out that the origin of the annual periodicity is still an open issue. Therefore, we need other Earth and spacecraft data to further confirm our inference. 

The Rieger-type periodicity of 157.7 days is also extracted from the two indices (IM11), and their information and waveforms are listed in Table 1 and shown in Figure 3. The mode exists in solar cycles 21 to 23, as well as in the ascending phase of cycle 24. However, it should be noted that the amplitude in cycle 21 is significantly larger than that in cycles 22 and 23, and the smallest amplitude is cycle 24, regardless of whether it is in the MPSI or the MWSI. Furthermore, we found that the intermittent amplitude variation of IM11 seems to deviate from the location of the peak of the 11 year solar cycle, and prefers to appear in the descending phase of cycle 21, and in the ascending phase of cycle 22 and 23. Figure 5 shows this phenomenon, where the plot of the Rieger-type periodicity and that of its corresponding 11 year periodicity are superposed. \cite{2002SoPh..205..177R} analyzed the coronal index of solar activity, and found that the spectral power of the 150 day period suddenly enhances before and after magnetic activity maximum. The result of our analysis is in good agreement with their finding.

The Rieger-type periodicity has been found in some solar activity indices, such as the X-flare index \citep{2008MNRAS.386.2278D}, sunspot group numbers \citep{1999ApJ...522L.153B}, and solar type III radio bursts \citep{2012ApJ...754L..28L}. The indices are related to strong magnetic field activities on the solar surface. \cite{1989ApJ...337..568L} found that the Rieger-type periodicity is present in the sunspot blocking function and 10.7 cm radio flux, but is absent in the plage index daily data. Based on this, they suggested that the periodicity should be generated by strong magnetic field activities.

Comparing with the relative strength ratio of IM11 in Table 1, the mean strength ratio of the Rieger-type periodicity in MWSI is larger than that in the MPSI. Because the MPSI and the MWSI reflect weak and strong magnetic field activities, respectively, it is easier to explain why the mode is more easily found in the strong magnetic indices. This finding also demonstrates that the analysis mode based on the SST is suitable for extracting those low-power signals. This reveals that it is not only associated with the strong magnetic fields, but is also related with the weak fields. In other words, the Rieger-type periodicity is a global fundamental phenomenon of the whole Sun. 

However, the origin of the Rieger-type periodicity is still not a satisfactory explanation. Because the timescale is close to the multiple of 27 days, some authors \citep{1987Natur.327..601B,2003ApJ...591..406B} speculated that it is a subharmonic oscillation caused by a 27 day signal, which approximates the sidereal rotation periodicity of the Sun near the equator. However, \cite{2010ApJ...709..749Z} and \cite{2000ApJ...540.1102L} supposed that the Rieger-type periodicity is related to the physical feature of the Rossby waves or the $r$-mode. The unstable harmonics of magnetic Rossby waves lead to a periodic emergence of magnetic flux on the solar surface. Moreover, the growth rates of the Rossby waves are sensitive to the magnetic field strengths.

\cite{2000ApJ...540.1102L} defined an equation to calculate the periodicities generated by the Rossby waves. The definition about the equation is as follows:
\begin{equation}
P_r\cong 25.1[m/2+0.17(2n+1)/m]. 
\end{equation}

Here, we use it to evaluate the extracted periodicities from the two indices. Assuming $n=1$, the significant periodicities of the Rossby waves are obtained: $\sim$102 days (m=8), $\sim$115 days (m=9), $\sim$152 days (m=12), $\sim$251 days (m=20), and $\sim$276 days (m=22). The periodicities of IM12, IM11, and IM10 listed in Table 1 are in better agreement with the Rossby wave periodicities, especially the periodicities of 251.6 and 276.2 days in IM10. So, we prefer to believe that the periodicities, including the Rieger-type periodicity, are related to the Rossby waves.
 
\begin{figure}[t]
	\centering
        \includegraphics[width=8.5cm]{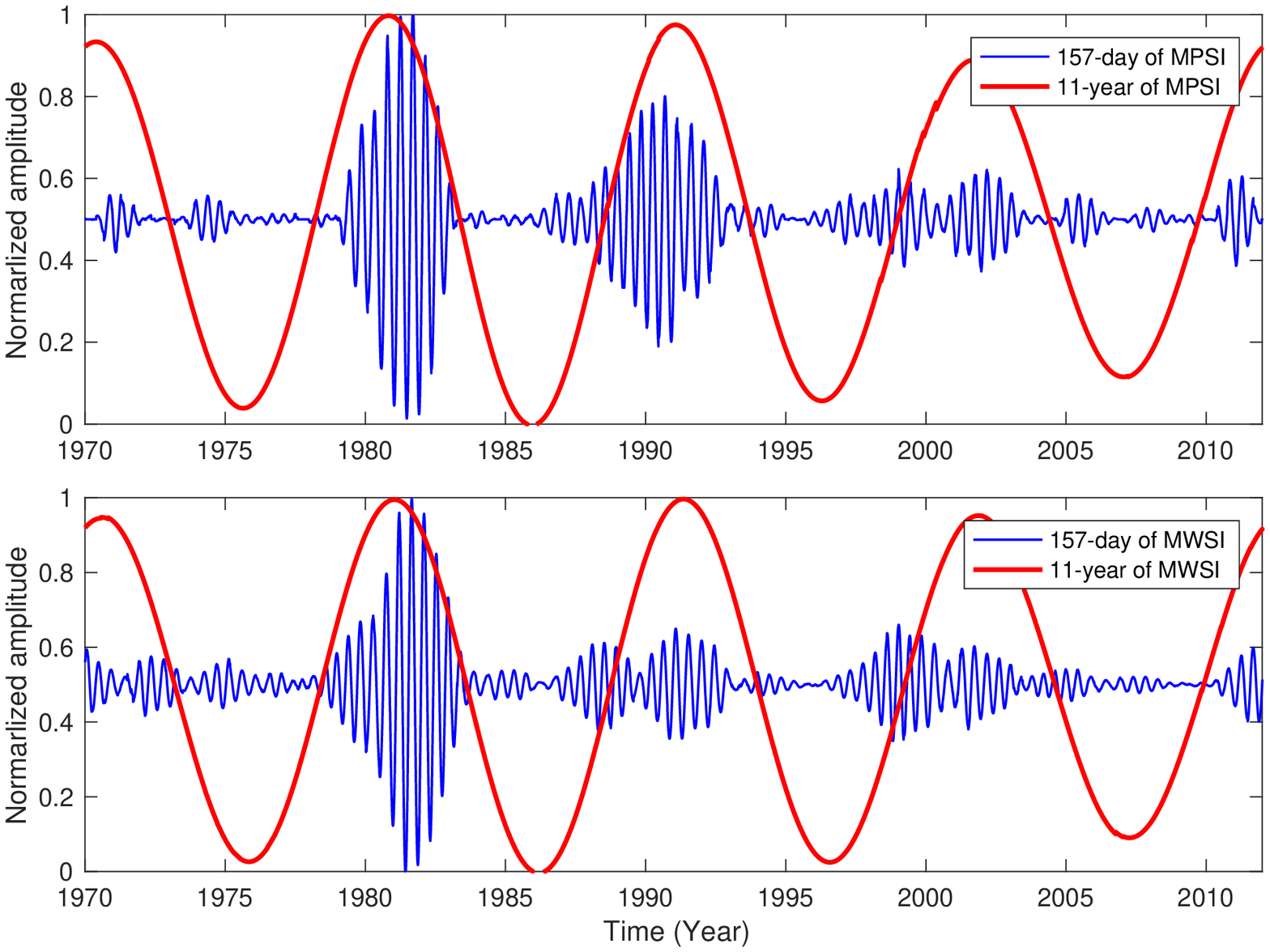}
        \caption{Upper panel: the Rieger-type periodicity of 157.7 days of the MPSI plotted with the blue curve, and the corresponding 10.6 year periodicity with the red curve. Bottom panel: the periodicities of 157.7 days and 10.6 years of the MWSI. All modes are normalized to their maximum values for better visualization.}
        \label{fig5}
\end{figure}

In our study, the periodicities of $\sim$1.3 and 1.7 years are also found. The 3.37 and 3.30 year periodicities (IM4) are seen to be approximately twice that of the 1.7 and 1.67 year (IM7) periodicities of the MPSI and the MWSI. Similarly, the 2.50 and 2.43 year periodicities (IM5) are inferred to be two multiple harmonics of the 1.3 and 1.33 year (IM8) periodicities. The periodicities found imply that they exist on the solar surface, regardless of the strong and weak magnetic field regions. \cite{2000Sci...287.2456H} determined a $\sim$1.3 year periodicity in the solar internal rotation rate near the base of the convection zone, from the \textit{Solar and Heliospheric Observatory}/Michelson Doppler Imager and the Global Oscillation Network Group data observed from 1995 May to 1999 November; this periodicity was absent after 2000 \citep{2007AdSpR..40..915H}. But, they also further emphasized that the phenomenon could have been caused by the intermittent periodic variation. \cite{2013ApJ...778...28C} found the periodicity in the soft and hard X-ray emission after 2001.

\begin{figure}[t]
	\centering
        \includegraphics[width=8.3cm]{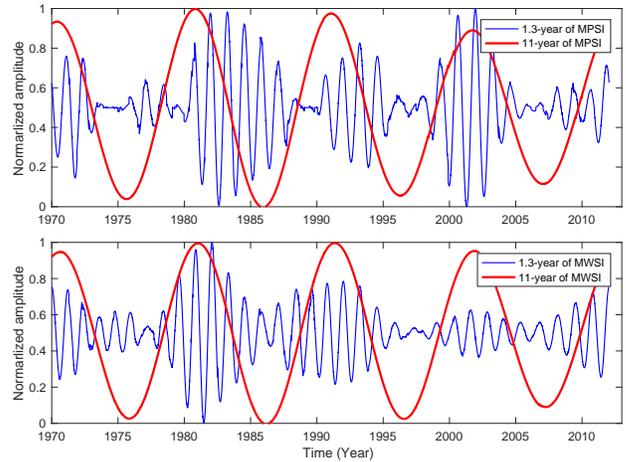}
        \caption{Upper panel: the 1.3 year periodicity in the MPSI plotted by a blue curve, and the corresponding 10.6 year periodicity plotted with a red curve. Bottom panel: the 1.33 year and 10.6 year periodicites of the MWSI. All modes are normalized to their maximum values for better visualization.}
        \label{fig6}
\end{figure}

To better illustrate the phase relation between the $\sim$11 year and $\sim$1.3 year periodicities, they were superposed on each other, and are shown in Figure 6. One can see that the amplitude of 1.3 year periodicity in the MWSI exhibits intermittent variations from the 11 year peak of the solar activity cycle (in the bottom panel of Figure 6), and slight deviations in the MPSI (shown in the upper panel of Figure 6). We also note that the periodic signal is still present after 2001.   

The periodicities of 1.3 and 1.7 years have been found from the interior of the Sun to the solar atmosphere, as well as from the interplanetary space to Earth. \cite{2014SoPh..289..707C} noted that the coherent relation between the interplanetary space and the Earth, which was not present in the Sun. \cite{2015JASTP.122...18D} inferred the scenario about the periodicity that the periodic energy emerges from the base of the convection zone into the solar atmosphere. Some of the energy are dissipated in the low atmosphere, and some of it eventually escape from the Sun in the form of solar wind and coronal mass ejections. So, the periodicities can be found in the interplanetary space and the Earth. Because the MPSI and the MWSI reflect the magnetic field variations of the solar surface, the energy of the periodicity transferred from the interior of the Sun to the solar surface causes the fluctuations of the magnetic field strengths. It notes that the dissipation process about dissipation needs to obtain the phase relation among the solar activity indices in different atmospheric layers and temporal observation data.

\section{Conclusion}
In this paper, we used the magnetic indices, the MPSI and the MWSI, measured by the 150 foot solar tower of the MWO since 1970 January 19, to investigate the quasi-periodic fluctuations of solar magnetic field strengths. The MPSI and the MWSI are considered as the weak and strong magnetic field variations on the solar surface. To accurately extract the periodicity modes of the two indices, we used a novel time--frequency analysis technique, called the SST. Our analysis reveals that the solar magnetic indices exist in multiple periodic modes, including those with the shortest periodicity of $\sim$13 days and the longest periodicity of $\sim$11 years, and in the midterm with timescale between $\sim$27 days and $\sim$11 years. Our main findings are as follows:

(1) The Rieger-type periodicity is extracted from the MPSI and the MWSI, demonstrating that the Rieger periodicity is a global phenomenon. Apart from the 157.7 day periodicity, the modes with $\sim$108, $\sim$116, $\sim$251, and $\sim$276 days are very close to those periodicity generated by the Rossby waves theoretical model. So, we infer the modes are modulated or controlled by the Rossby waves or $r$-mode oscillations.

(2) The MPSI and the MWSI represent the magnetic field variations of the solar surface, and our analysis result indicates that the timescales with 1.3 and 1.7 years not only exist in the cycles 21, 22, and 23, but also in the ascending phase of cycle 24. Based on this, we deduce that the periodicity is significant periodic modes on the solar surface. The periodicities have been found from in the interior and high atmospheric layer to the interplanetary space and the Earth. So, its origin is suggested from the interior of the Sun, and its energy transfers toward the solar surface. It then dissipates to high astomspheric layer. Our finding just fills the link about the solar surface. 
  
Finally, it is particularly pointed out that our finding benefits from the ability of the SST in terms of suppressing spectrum smearing and mode mixing. Hence, it is suitable for representing and decomposing those intrinsic mode in a non-stationary signal, for example, solar quasi-periodic fluctuations. 

\acknowledgments
We would like to express our gratitude to the anonymous referee for the careful reading of the paper and constructive comments that improved our paper. This study includes the MPSI and MWSI data from the synoptic program at the 150 foot Solar Tower of the Mount Wilson Observatory, which is operated by UCLA, with funding from NASA, ONR, and NSF, under agreement with the Mt. Wilson Institute. This work is supported by the National Natural Science Foundation of China (Numbers 11463003, U1531132, and U1631129), the National Key Research and Development Program of China (Number 2016YFE0100300), and the CAS Key Laboratory of Solar Activity of National Astronomical Observatories (KLSA201715).

\bibliographystyle{aasjournal}
\bibliography{sst}


\end{document}